\documentclass[twocolumn]{aastex631}

\usepackage{amsmath}
\usepackage{graphicx}
\usepackage{textcomp}
\usepackage{gensymb}
\usepackage{natbib}
\usepackage{longtable}
\usepackage{multirow}
\usepackage{float}
\usepackage{enumitem}

\begin{document}
\def\teff{$T_{\rm eff}$}
\def\cs{$\chi^{2}$}
\def\rsun{$R_{\odot}$}
\def\msun{$M_{\odot}$}
\def\rstar{$R_{\star}$}
\def\rearth{$R_{\earth}$}
\def\av{$A_{V}$}
\def\emcee{\texttt{emcee}}
\def\kepler{\textit{Kepler}}

\graphicspath{{figures/}}
\shorttitle{Binary Planet Hosts III.}
\shortauthors{Sullivan et al.}

\title{Revising Properties of Planet-Host Binary Systems. III. There is No Observed Radius Gap For \kepler\ Planets in Binary Star Systems\footnote{Based on observations obtained with the Hobby-Eberly Telescope, which is a joint project of the University of Texas at Austin, the Pennsylvania State University, Ludwig-Maximilians-Universität München, and Georg-August-Universität Göttingen.}}

\author[0000-0001-6873-8501]{Kendall Sullivan}
\altaffiliation{NSF Graduate Research Fellow}
\affil{Department of Astronomy, University of Texas at Austin, Austin, TX, 78712, USA}

\author{Adam L. Kraus}
\affil{Department of Astronomy, University of Texas at Austin, Austin, TX, 78712, USA}

\author[0000-0001-8832-4488]{Daniel Huber}
\affiliation{Institute for Astronomy, University of Hawai`i, 2680 Woodlawn Drive, Honolulu, HI 96822, USA}

\author[0000-0003-0967-2893]{Erik A. Petigura}
\affil{Department of Physics \& Astronomy, University of California Los Angeles, Los Angeles, CA, 90095, USA}

\author{Elise Evans}
\affil{Institute for Astronomy, University of Edinburgh, Royal Observatory, Blackford Hill, Edinburgh, EH9 3HJ, UK}

\author[0000-0001-9823-1445]{Trent Dupuy}
\affil{Institute for Astronomy, University of Edinburgh, Royal Observatory, Blackford Hill, Edinburgh, EH9 3HJ, UK}

\author{Jingwen Zhang}
\altaffiliation{NASA FINESST Fellow}
\affiliation{Institute for Astronomy, University of Hawai`i, 2680 Woodlawn Drive, Honolulu, HI 96822, USA}

\author[0000-0002-2580-3614]{Travis A. Berger}
\altaffiliation{NASA Postdoctoral Program Fellow}
\affil{Exoplanets and Stellar Astrophysics Laboratory, Code 667, NASA Goddard Space Flight Center, Greenbelt, MD 20771, USA}

\author[0000-0002-5258-6846]{Eric Gaidos}
\affil{Department of Earth Sciences, University of Hawai’i at M\"{a}noa, Honolulu, HI, 96822, USA}

\author[0000-0003-3654-1602]{Andrew W. Mann}
\affil{Department of Physics and Astronomy, The University of North Carolina at Chapel Hill, Chapel Hill, NC, 27599, USA}

\correspondingauthor{Kendall Sullivan}
\email{kendallsullivan@utexas.edu}

\begin{abstract}
Binary stars are ubiquitous; the majority of solar-type stars exist in binaries. Exoplanet occurrence rate is suppressed in binaries, but some multiples do still host planets. Binaries cause observational biases in planet parameters, with undetected multiplicity causing transiting planets to appear smaller than they truly are. We have analyzed the properties of a sample of 119 planet-host binary stars from the \kepler\ mission to study the underlying population of planets in binaries that fall in and around the radius valley, which is a demographic feature in period-radius space that marks the transition from predominantly rocky to predominantly gaseous planets. We found no statistically significant evidence for a radius gap for our sample of 122 planets in binaries when assuming the primary stars are the planet hosts, with a low probability ($p < 0.05$) of the binary planet sample radius distribution being consistent with the single-star small planet population via an Anderson-Darling test. These results reveal demographic differences in the planet size distribution between planets in binary and single stars for the first time, showing that stellar multiplicity may fundamentally alter the planet formation process. A larger sample and further assessment of circumprimary versus circumsecondary transits is needed to either validate this non-detection or explore other scenarios, such as a radius gap with a location that is dependent on binary separation.
\end{abstract}


\section{Introduction}\label{sec:intro}
The \kepler\ mission \citep{Borucki2010} discovered most of the known small planets, and the \kepler\ catalog of confirmed and candidate planets \citep[e.g.,][]{Mullally2015, Coughlin2016, Thompson2018} remains a powerful source for the demographics of those small planets. The combination of high-precision stellar characterization from efforts like the California \kepler\ Survey (CKS; \citealt{Petigura2017, Johnson2017}) or Gaia \citep{Gaia2016} and the \kepler\ planet catalog has increased the achievable precision of planet host star parameters \citep[e.g.,][]{Berger2020}. Precise stellar parameters have led to the discovery or confirmation of dramatic features in the \kepler\ exoplanet population, such as the sub-Neptune desert (a dearth of short-period Neptune-size planets; \citealt{Mazeh2016}) and the radius gap \citep[e.g.,][]{Fulton2017}.

The radius gap (or radius valley) is a feature of exoplanet populations that separates super-Earths from sub-Neptunes, first suggested by \citet{Owen2013}, observed by \citet{Fulton2017} in the \kepler\ sample, and subsequently studied by others \citep[e.g.,][]{VanEylen2018, Berger2020, Cloutier2020, Hardegree-Ullman2020, David2021, Hansen2021, Petigura2022}. The radius gap is a dearth of planets with radii of $1.5 < R_{\oplus} < 2$, centered at 1.8-1.9 \rearth\ \citep[although the location and slope of the gap are both dependent on the mass, metallicity, and age of the sample; ][]{Hardegree-Ullman2020, Petigura2022}, and it is typically interpreted as the gap between rocky planets and those with a substantial hydrogen/helium atmosphere. Alternatively, it may demarcate a separation between rocky planets and water worlds \citep{Luque2022}. If a rocky planet retains even a small H/He envelope, the large scale height of the gas causes a rapid increase in radius, creating a gap. The super-Earths that fall near the radius gap may be highly-irradiated cores of former gas giants \citep[e.g.,][]{Owen2013, Berger2020}, the outcome of core-powered mass loss \citep[e.g.,][]{Ginzburg2018, Gupta2020} or the rare cases where a thin atmosphere is all that formed because of scenarios like gas-poor planet formation \citep{Lee2021, Lee2022}. 

\kepler's discoveries combined with high-precision follow-up have revealed planetary demographic features in detail, but the population of planets in binary stars remains largely unexplored despite many \kepler\ Objects of Interest (KOIs) occurring in binary systems. Although the \kepler\ Input Catalog (KIC; \citealt{Brown2011}) construction process selected against binaries that were identified in ground-based preliminary imaging, additional close binaries in the KOIs were discovered after the fact in follow-up observations \citep[e.g.,][]{Law2014, Kraus2016, Atkinson2017, Furlan2017, Ziegler2017}. The presence of an additional star in a system, especially at separations of $\rho < 50$ au, can impact planet formation and survival \citep[e.g.,][]{Cieza2009, Harris2012, Kraus2012, Kraus2016, Zurlo2020, Moe2021}. A binary within $\sim 2-3\arcsec$ (approximately the average seeing during the KIC observations and half the size of a \kepler\ pixel; \citealt{Borucki2010, Brown2011}) can impact measured planet parameters \citep{Furlan2020, Sullivan2022b, Sullivan2022c}, leading to inferred planet radii that are too small and equilibrium temperatures that are too low. 

The observational effects of multiplicity obscure differences in properties of planets in single and binary stars, biasing the results of single-star demographics and limiting our knowledge of planet demographics in binary stars. Because circumstellar disks in close binary systems have reduced lifetimes and masses \citep{Cieza2009, Harris2012, Kraus2012, Zurlo2020}, the properties of the planets in those systems may be impacted. One example is the decreased occurrence rate of planets as a function of separation \citep[e.g.,][]{Wang2014, Wang2015, Kraus2016, Ziegler2020, Moe2021}. Another possible impact of multiplicity is that the reduced circumstellar disk mass and lifetime could cause a lower average core mass and radius for rocky planets, shifting the radius gap to smaller planetary radii. The location of that shift could be dependent on the binary mass ratio and separation, where a closer binary or more massive secondary would be more disruptive to planet formation and thus cause smaller cores and a gap that occurred at smaller planetary radii.

To investigate the properties of the radius gap in binary stars, we developed an algorithm to accurately retrieve the properties of close binary stars ($\rho < 2\arcsec$) by simultaneously fitting a composite low-resolution red-optical spectrum of the system, unresolved catalog photometry, and speckle and adaptive optics (AO) contrasts \citep{Sullivan2022b}. We used the red-optical low-resolution spectrograph LRS2-R on the Hobby-Eberly Telescope (HET) at McDonald Observatory to observe a sample of 119 stars hosting 122 small planets ($R < 2.5$ \rearth). We retrieved the component stellar parameters (\teff\ and radius) and used them to revise the planetary radii, which we used to explore the properties of the radius gap in binary star systems.

\section{Data Collection}\label{sec:data}
\subsection{Sample Selection} \label{subsec:sample}
To select a sample of super-Earths and sub-Neptunes spanning the radius gap, we queried the IPAC Exoplanet Archive \citep{exoplanetarchive} on 20210201 for \kepler\ Objects of Interest (KOIs) that hosted at least one planet with radius $R_{p} \leq 2.5$ \rearth\ and a disposition of ``Confirmed Planet'' or ``Planet Candidate''. We chose a radius cutoff of 2.5 \rearth\ for our sample because we were initially investigating super-Earths in binary stars. Future work will expand the sample to larger radii to analyze a larger number of sub-Neptunes.

We cross-matched that sample with the high-resolution imaging compilation of \citet{Furlan2017} to identify binary systems, which we then restricted to systems with one stellar companion with separation $\rho < 2 \arcsec$ and no other companions within $3.5 \arcsec$ and at least one measured stellar component contrast $\Delta m_{[\lambda]} \leq 3.5$ mag. These criteria were set to ensure that at least one planet in each system fell near the radius valley and that the secondary component of the binary star was both close and bright enough to impact the derived \kepler\ planetary parameters. We excluded triples and higher-order multiples because the solution was too degenerate to retrieve good stellar parameters for the tertiary.

\begin{deluxetable}{cc}
\tablecaption{Excluded Systems \label{tab:excluded}}
\setcounter{table}{0}
\tablecolumns{2}
\tablehead{\colhead{KOI} & \colhead{Reason for exclusion}}
\startdata
KOI-0126 &	Quad\\
KOI-0279 &	Triple\\
KOI-0307 &	Quad\\
KOI-0353 &	Triple\\
KOI-0364 &	Quad\\
\enddata
\tablecomments{The full table is available from the supplementary material.}
\end{deluxetable}

\begin{figure}
    \plotone{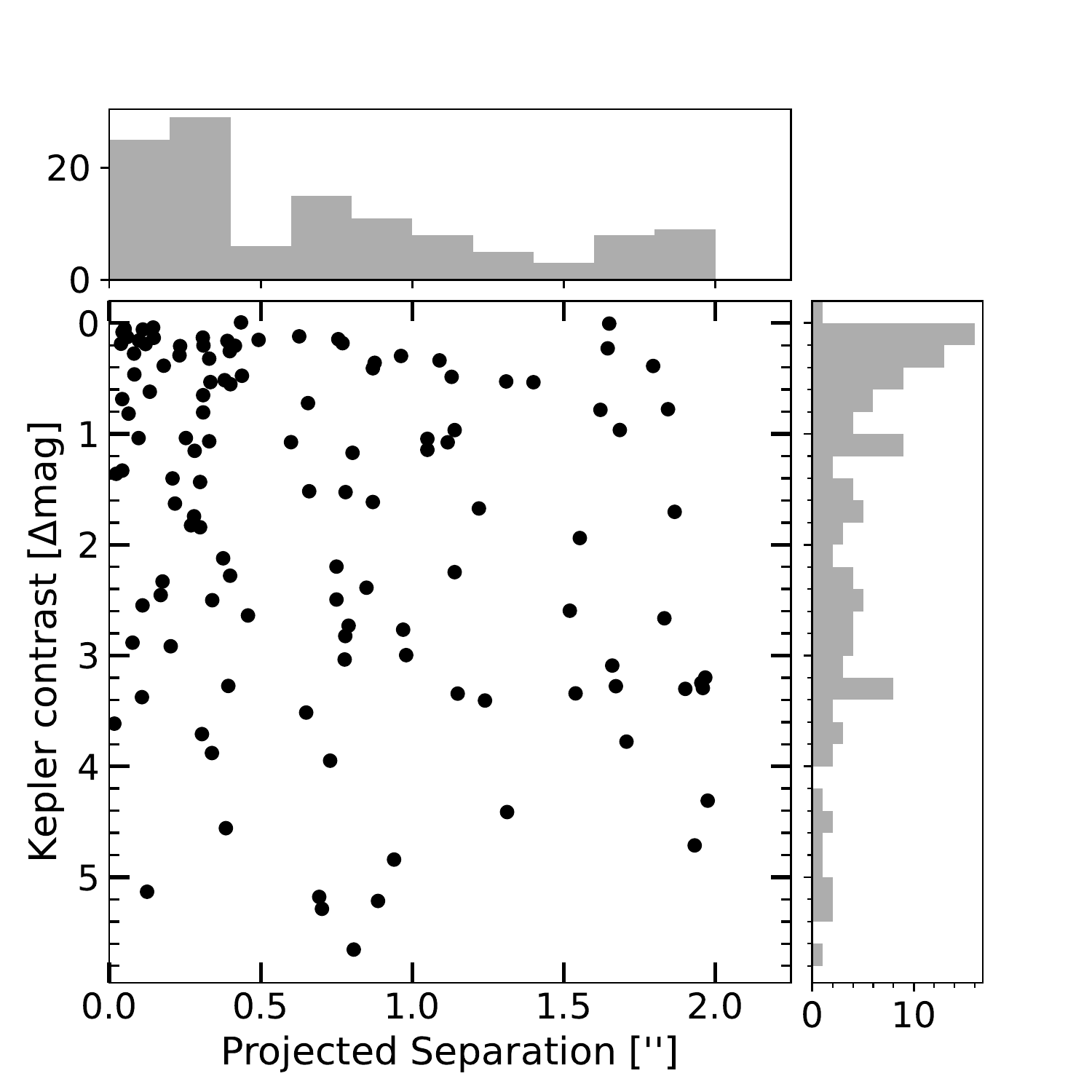}
    \caption{The separation and \kepler\ contrast for our observed sample. Although we had a maximum contrast cutoff of 3.5 mag in at least one photometric band, this led to the inclusion of some systems with high optical contrasts but with a NIR contrast that met our selection criterion. The sample only included systems out to angular separations of 2$\arcsec$. Although there is substantial scatter in both dimensions, the most frequent systems are those at close separation and near-equal brightness. The histograms at the top and on the right of the figure show the marginalized distribution in contrast or separation.}
    \label{fig:input_vals}
\end{figure}

After assembling the initial target list, we visually inspected any available high-contrast imaging data on the ExoFOP \citep{exofop} and the Keck Observatory Archive%
\footnote{\url{koa.ipac.caltech.edu/}}
for each target to ensure that it was a binary. Almost all of our targets had some imaging available for inspection. Following an inspection, we removed any targets that did not have an apparent secondary in any of the available images, or targets that appeared to have a higher-order multiple component. After removing the 70 anomalous or high-order multiple systems, we had 133 targets. Because our observations occurred in queue mode, we observed a total of 122/133 systems to achieve 90\% completeness. Table \ref{tab:excluded} lists the excluded systems with our rationale for exclusion. Figure \ref{fig:input_vals} shows the separation and \kepler\ contrast distributions for our sample.

Table \ref{tab:source_params} lists the targets in our sample, as well as the separation and literature contrasts for each system, which were mostly taken from \citet{Furlan2017}, who took their own observations as well as compiling results from a number of different works. Among these other surveys, \citet{Horch2012} used the DSSI instrument on Gemini North; \citet{Dressing2014} used the ARIES instrument on the MMT; \citet{Wang2015} used PHARO at Palomar Observatory and NIRC2 at Keck; \citet{Everett2015} used DSSI speckle imaging at Gemini North and NIR AO imaging at several different sites; \citet{Kraus2016} used NIRC2 at Keck Observatory; \citet{Baranec2016} used Robo-AO on the Palomar Observatory 1.5m telescope; \citet{Atkinson2017} used NIRC2 on Keck; \citet{Ziegler2017} and \citet{Ziegler2018} used Robo-AO on the Palomar 1.5m, Gemini NIRI LGS-AO, and Keck NIRC2 LGS-AO.

Some systems had only one contrast reported in \citet{Furlan2017}. In some cases, the components of the binary were resolved by the {\it Gaia} mission in EDR3, so we were able to include a $\Delta m_{G}$ with the other contrast to ensure a good constraint on the contrast fit. Of the 119 systems, 36 were resolved in {\it Gaia} EDR3. For several systems (KOI-1972, KOI-1973, KOI-2962, KOI-3073, KOI-3234, KOI-4823, KOI-5845, KOI-5971, and KOI-6475) we had additional high-contrast $K$-band NIRC2 imaging from our group that we included in the fits and report in Table \ref{tab:source_params}. The observing strategy and data reduction for those systems match those described in \citet{Kraus2016}. 

We also revised the properties of the remaining systems with only one contrast, but there may be larger systematic errors in those results. However, we do not expect the systematic error to significantly impact the results or conclusions of our analysis. Most of the systems with only one contrast had a $\Delta m_{K}$. We ran a test on other systems with multiple contrasts by analyzing them with only the $\Delta m_{K}$ versus the NIR plus speckle contrasts and found that the derived stellar parameters did not change significantly, meaning that all but five of the systems with one contrast likely have minimal error introduced because of only having a single contrast.

For several systems, we found contrasts in similar filters that had inconsistent values (e.g., did not match even though the observations were taken in filters with similar wavelengths). In general, we chose to keep the contrasts that were more consistent with the underlying contrast curve and/or the other measurements available for the system. The excluded contrasts are listed in Table \ref{tab:source_params} for completeness, and are listed here, but were excluded from subsequent analysis. KOI-1531 had a $\Delta m_{LP600}$ that was inconsistent with both other contrasts, so we removed it from the analysis. KOI-1630 had an inconsistent $\Delta m_{J}$. KOI-2067 had a $\Delta m_{J}$ that was inconsistent with all other contrasts. KOI-2413 had a $\Delta m_{LP600}$ that was inconsistent. KOI-2657 had an inconsistent $\Delta m_{[692 nm]}$ measurement. KOI-3069 had an inconsistent $\Delta m_{K}$. KOI-4287 had a $\Delta_{LP600}$ that was inconsistent. 

KOI-3112 and KOI-6513 are new candidate triple systems because their fits predict a larger, cooler primary star and a smaller, hotter secondary, which would be consistent with the primary being an unresolved multiple. However, we analyzed them as binary stars because the possible tertiary was not resolved and so did not have literature contrasts.

\subsection{Observations with LRS2 on the HET}\label{subsec:obs}
We observed all systems using the red setting (6500 $\leq \lambda \leq $ 10500 \AA; R $\sim$ 1800) of the second-generation low-resolution spectrograph (LRS2-R; \citealt{Chonis2014, Chonis2016}) on the HET at McDonald Observatory. Our observations were taken between UT dates 20210402 and 20210920 in queue observing mode under spectroscopic conditions during bright or gray time. Because the observations were taken in queue mode, our sample does not include every target fitting our selection criteria because of the random nature of queue scheduling. However, our sample was observed to $\sim 90\%$ completeness, and any targets not included were excluded randomly based on the queue scheduling, so we do not expect that those systems would change our scientific conclusions.

The instrument details and our observing strategy are described in \citet{Sullivan2022c}. Briefly, we observed each target for either 300s (which was comparable to the typical overhead from target acquisition) or the time required to reach a S/N of 100, whichever was larger. Our typical integration times were 300-600s. Because our sources did not need to be resolved and were typically fairly bright, we set a high seeing threshold of 2.5\arcsec. This was also feasible because of the instrumental design of LRS2, which is an integral field spectrograph (IFS) with a 12\arcsec $\times$ 6\arcsec\ field of view that is continuously tiled by hexagonal 0\arcsec.6 lenslets. Because the instrument is an IFS with no slit losses, high-quality data extraction remained possible in observations that were taken in poor seeing.

After data reduction, the source was extracted using an aperture clipped at 2.5 times the seeing, calculated in the wavelength frame with the highest S/N. Although the red setting of LRS2 (LRS2-R) observes in two arms, red (6500 $< \lambda < $ 8470 \AA) and far-red (8230 $< \lambda < $ 10500 \AA), low S/N and telluric contamination reduced the quality of the data from the far-red arm, so our analysis was restricted to only the data from the red arm.

\section{Stellar Parameter Retrieval}
After collecting observations for our targets, we removed the atmospheric absorption, then retrieved the stellar parameters (\teff, \av, radius, and radius ratio) using an MCMC fitting algorithm. The details of most of our stellar analysis methods are described in \citet{Sullivan2022b} and \citet{Sullivan2022c} but we briefly describe them here, emphasizing any changes between those papers and this analysis.

To correct for the telluric absorption, we used \texttt{scipy optimize} to fit for the best humidity using a composite spectrum created using the BT-Settl stellar atmosphere models \citep{Allard2013, Rajpurohit2013, Allard2014, Baraffe2015} with the \citet{Caffau2011} line list, combined with a grid of model telluric spectra generated from TelFit \citep{Gullikson2014}. We optimized the humidity by simultaneously fitting the humidity and a composite best-fit temperature using an initial guess consistent with the \kepler\ DR25 stellar temperature \citep{Mathur2017}. We divided out the best-fit telluric spectrum, then masked the regions with the most substantial residuals: two large oxygen absorption bands and a strong line in a water feature at $6860-6880$ \AA, $7600-7660$ \AA, and $8210-8240$ \AA, respectively. 

After removing the telluric features, we assumed that each binary was parameterized by a set of values $\theta =$ \{\teff$_{1}$, \teff$_{2}$, \av, $R_{1}$, $R_{2}/R_{1}, [\pi]$\}, where parallax ($\pi$) was only included for systems where {\it Gaia} parallaxes were available, and $R_{1}$ and $R_{2}/R_{1}$ were the primary star radius and the radius ratio between the primary and secondary star. For 17\% of our systems, a {\it Gaia} EDR3 parallax was unavailable, so we used the distance listed on ExoFOP from \citet{Mathur2017} as the distance prior with an inflated error (a 10\% error in parallax) to account for the reduced accuracy of those measurements. For systems where we used the \citet{Mathur2017} distance, we did not implement a distance prior, but only used the distance to inform the initialization of the radius guess. Unlike our past work, we did not fit for surface gravity, because we found in \citet{Sullivan2022b} and \citet{Sullivan2022c} that we were unable to place constraints on surface gravity because of the low resolution of our spectroscopic observations. Instead, we calculated surface gravity for a given \teff\ using the MESA Isochrones and Stellar Tracks (MIST) stellar evolutionary models \citep{Paxton2011, Paxton2013, Paxton2015, Dotter2016, Choi2016} at an age of 1 Gyr and used that surface gravity when querying the spectral models. The choice of 1 Gyr of age was arbitrary and served to ensure that our inferred system radii fell on the main sequence. However, our fitting method was able to identify possible interlopers with radii that were inconsistent with two stars on the main sequence by producing nonphysical temperatures and radius ratios, even with the radius constraint, so our results were ultimately agnostic to our assumed age. 

We assumed solar metallicity for all systems. Metallicity does change stellar radius slightly, as well as possibly impacting the properties of sub-Neptunes above the radius gap \citep{Chen2022}, but our spectra did not cover typical lines (e.g., Ca H and K or the Ca IR triplet) that can be used to measure metallicity at this spectral resolution \citep{Lee2008}. In addition, we ran tests using model spectra of different metallicities and found that our results did not change significantly. \citet{Berger2020} found that the median metallicity of the \kepler\ sample was about solar, but with some spread to $\pm$0.5 dex. However, a metallicity change of 0.5 dex in either the positive or negative direction represents a relatively small change in stellar radius (a $\sim$3\% change for $\pm$0.5 dex compared to solar metallicity), which is much smaller than the average planetary radius change of 16\% for a primary star, and so should only have a small effect on the total error on the stellar radius. Therefore, we note that the assumption of solar metallicity may slightly change the planetary radii we derive, but should not significantly impact our conclusions.

Using the BT-Settl models\footnote{\url{https://phoenix.ens-lyon.fr/Grids/BT-Settl/CIFIST2011/}}, we created synthetic test data sets of unresolved photometry in the KIC filters \citep{Brown2011}, contrasts in the appropriate filters of the literature contrasts, and a low-resolution unresolved binary red-optical spectrum to match the observed data set of KIC photometry, literature contrasts, and the HET/LRS2-R spectrum. Because of flux calibration uncertainty in the spectrum, we continuum-normalized the LRS2-R data using a low-order polynomial fit. In practice, the consequence of this choice was that the temperature was jointly constrained by the spectral lines and the shape of the SED described by the unresolved catalog photometry, while the extinction was constrained by the $A_{V}$ prior and the shape of the SED. We fit the combined data set to find the best parameters using a Gibbs sampler that was modified to only take steps downward in \cs\ space, instead of being able to randomly choose to step upward in \cs\ space. Following the optimization step, we assessed the best-fit parameters and their statistical errors using \emcee\ \citep{Foreman-Mackey2013} initialized with the best 30\% of walkers from the optimization stage. 

During each stage, we imposed several priors. For temperature, we imposed a uniform prior between $3000 < T_{\rm eff} < 7000$ K. For $R_{1}$ and $R_{2}/R_{1}$ we imposed a Gaussian radius prior with a mean value inferred from the \teff\ guess using the MIST evolutionary models at an age of 1 Gyr and at solar metallicity. This was an arbitrary age choice representing a time when the full mass range of our sample was on the main sequence, and should not affect our final conclusions. We used the radius prior to ensure that both stars would be consistent with being drawn from the main sequence. The standard deviation of the Gaussian prior on the radius was 5\%, which is close to the precision of typical radius measurements \citep{Mann2015, Tayar2022}. If a {\it Gaia} parallax was available for the system, we imposed a Gaussian prior in parallax space using the value and error of the {\it Gaia} parallax as the mean and standard deviation of the Gaussian. If a {\it Gaia} parallax was not available, we used the inverted distance listed on ExoFOP and taken from \citet{Mathur2017}, assuming a Gaussian parallax prior with a width defined by the larger parallax error measured when inverting the distance and its error. For extinction, we imposed a Gaussian prior using the 3D Bayesian dust map \texttt{bayestar} \citep{Green2019} implemented in the \texttt{dustmaps} package\footnote{\url{https://dustmaps.readthedocs.io/en/latest/index.html}}, which has been shown to yield reasonably accurate extinction values in the \kepler\ field \citep{Rodrigues2014}. We used the mean and standard deviation of the samples at the appropriate location in 3D space as the parameters for the Gaussian extinction prior. 

\section{Revised System Parameters}

\begin{deluxetable*}{CCCCCCCCC}
\tablecaption{Stellar parameter fit results for all KOIs in our sample}\label{tab:star results}
\setcounter{table}{2}
\tablecolumns{9}
\tablewidth{0pt}
\tablehead{
\colhead{KOI} & \colhead{$T_{{\rm eff},1}$} & \colhead{$T_{{\rm eff},2}$} & \colhead{$T_{Kepler}$} & \colhead{$R_{1}$} & \colhead{$R_{2}/R_{1}$} & \colhead{$R_{Kepler}$} & \colhead{f$_{corr, p}$} & \colhead{f$_{corr, s}$}\\
\colhead{} & \colhead{(K)} & \colhead{(K)} & \colhead{(K)} & \colhead{\rsun} & \colhead{} & \colhead{\rsun} & \colhead{} & \colhead{}
}
\startdata
0042 & 6639$^{+14}_{-12}$ & 4771$^{+33}_{-37}$ & 6407$\pm$76 & 1.38$^{+0.01}_{-0.01}$ & 0.52$^{+0.01}_{-0.01}$ & 1.34 $\pm$ 0.07 & 1.06$^{+0.05}_{-0.05}$ & 2.27$^{+0.13}_{-0.11}$\\
0177 & 6091$^{+72}_{-117}$ & 5902$^{+82}_{-123}$ & 5706$\pm$114 & 1.05$^{+0.04}_{-0.06}$ & 0.93$^{+0.01}_{-0.01}$ & 1.03 $\pm$ 0.15 & 1.35$^{+0.22}_{-0.19}$ & 1.44$^{+0.24}_{-0.19}$\\
0191 & 5689$^{+46}_{-52}$ & 3969$^{+34}_{-33}$ & 5422$\pm$108 & 0.88$^{+0.01}_{-0.01}$ & 0.60$^{+0.01}_{-0.01}$ & 0.89 $\pm$ 0.09 & 1.02$^{+0.11}_{-0.10}$ & 2.79$^{+0.31}_{-0.26}$\\
0227 & 4140$^{+30}_{-28}$ & 4001$^{+12}_{-12}$ & 4094$\pm$81 & 0.59$^{+0.01}_{-0.02}$ & 1.02$^{+0.01}_{-0.01}$ & 0.57 $\pm$ 0.03 & 1.39$^{+0.08}_{-0.08}$ & 1.56$^{+0.10}_{-0.09}$\\
0270 & 5891$^{+82}_{-83}$ & 5860$^{+85}_{-73}$ & 5587$\pm$100 & 0.97$^{+0.04}_{-0.04}$ & 0.99$^{+0.01}_{-0.01}$ & 1.42 $\pm$ 0.05 & 0.96$^{+0.05}_{-0.05}$ & 0.96$^{+0.05}_{-0.05}$\\
\enddata
\tablecomments{The revised stellar temperatures and radii from this work, as well as the \kepler\ \citep{Mathur2017} values for the composite stellar system properties, and the planetary radius correction factor if the primary or secondary star is the host. A truncated version of the table is shown here for reference, and the remainder of the table is available in machine-readable format from the journal.}
\end{deluxetable*}

\begin{deluxetable*}{CCCCCCCCCC}
\tablecaption{Planet parameter fit results for all planets in the binary KOI sample}\label{tab:planet_results}
\setcounter{table}{3}
\tablecolumns{10}
\tablewidth{0pt}
\tablehead{
\colhead{KOI} & \colhead{R$_{p, pri}$} & \colhead{R$_{p, sec}$} & \colhead{R$_{Kep}$} & \colhead{T$_{eq,pri}$} & \colhead{T$_{eq,sec}$} & \colhead{T$_{eq, Kep}$} & \colhead{S$_{pri}$} & \colhead{S$_{sec}$} & \colhead{S$_{kep}$}\\
\colhead{} & \colhead{($R_{\earth}$)} & \colhead{($R_{\earth}$)} & \colhead{($R_{\earth}$)} & \colhead{(K)} & \colhead{(K)} & \colhead{(K)} & \colhead{(S$_{\earth}$)} & \colhead{(S$_{\earth}$)} & \colhead{(S$_{\earth}$)}
}
\startdata
42.01 & 2.58$^{+0.18}_{-0.18}$ & 5.53$^{+0.40}_{-0.41}$ & 2.43$\pm$0.12 & 911$^{+25}_{-26}$ & 470$^{+14}_{-14}$ & 866 & 145.44$^{+1.67}_{-3.10}$ & 15.60$^{+0.54}_{-0.71}$ & 132.94$\pm$20.74\\
177.01 & 2.44$^{+0.50}_{-0.52}$ & 2.62$^{+0.56}_{-0.56}$ & 1.78$\pm$0.26 & 722$^{+61}_{-59}$ & 677$^{+56}_{-56}$ & 669 & 56.47$^{+6.85}_{-7.08}$ & 44.56$^{+5.56}_{-5.63}$ & 47.39$\pm$19.64\\
191.01 & 11.19$^{+1.55}_{-1.63}$ & 30.68$^{+4.34}_{-4.28}$ & 10.89$\pm$1.11 & 695$^{+39}_{-40}$ & 377$^{+21}_{-21}$ & 662 & 52.64$^{+3.15}_{-3.23}$ & 6.52$^{+0.40}_{-0.39}$ & 45.46$\pm$14.22\\
191.02 & 2.33$^{+0.34}_{-0.35}$ & 6.35$^{+0.89}_{-0.94}$ & 2.25$\pm$0.23 & 1288$^{+71}_{-73}$ & 699$^{+39}_{-39}$ & 1226 & 619.10$^{+37.10}_{-37.99}$ & 76.62$^{+4.70}_{-4.58}$ & 535.84$\pm$167.62\\
191.03 & 1.23$^{+0.17}_{-0.18}$ & 3.38$^{+0.47}_{-0.49}$ & 1.20$\pm$0.12 & 1941$^{+104}_{-104}$ & 1053$^{+55}_{-57}$ & 1846 & 3181.11$^{+190.60}_{-195.22}$ & 393.71$^{+24.15}_{-23.54}$ & 2741.07$\pm$857.46\\
\enddata
\tablecomments{The revised and \citet{Thompson2018} planetary radii, instellations, and equilibrium temperatures. The entire sample of planets is included, although we only used systems with $R_{p} < 2.5$ \rearth\ for the remainder of the analysis. Five rows of the table are shown here for reference, and the full table is available in the supplementary material.}
\end{deluxetable*}

To revise the properties of close binary stars that host \kepler\ planets with radii in or below the radius valley (i.e., super-Earths), we developed an MCMC fitting algorithm to retrieve accurate parameters of spectroscopically unresolved binary stars. For each of the 119 stars in our sample, we analyzed a composite data set of unresolved spectra observed with HET/LRS2-R, unresolved catalog photometry from the KIC, and resolved contrasts from literature NIR AO or optical speckle imaging. Tables 3 and \ref{tab:planet_results} contain the revised parameters for all the stars and planets in our sample. The full tables are also available in machine-readable format from the journal. We removed any planets in a system that had original (unrevised) radii larger than 2.5 \rearth from our final analyss, but the revised planet parameters are still reported in Table \ref{tab:planet_results} for completeness.

\begin{figure*}
\plottwo{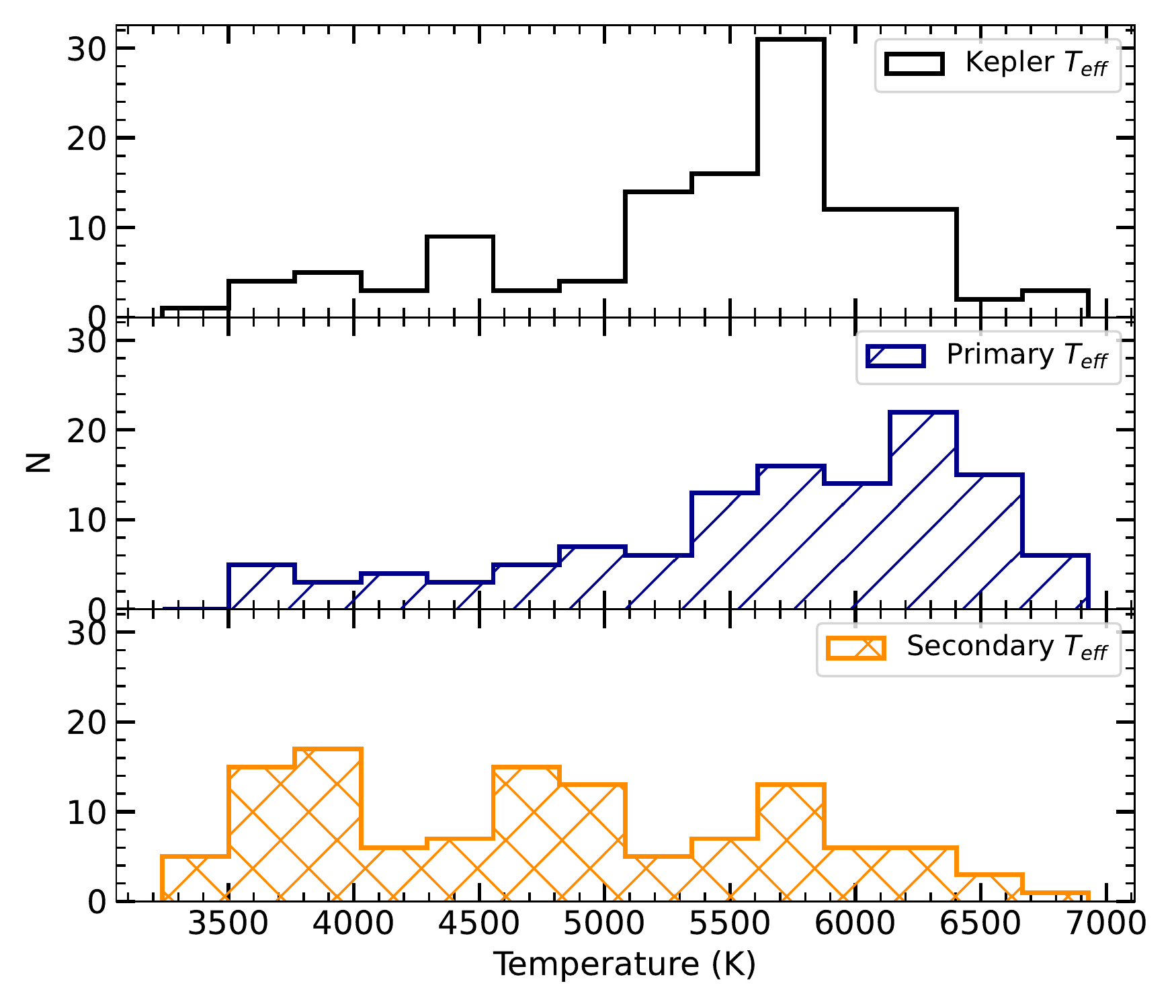}{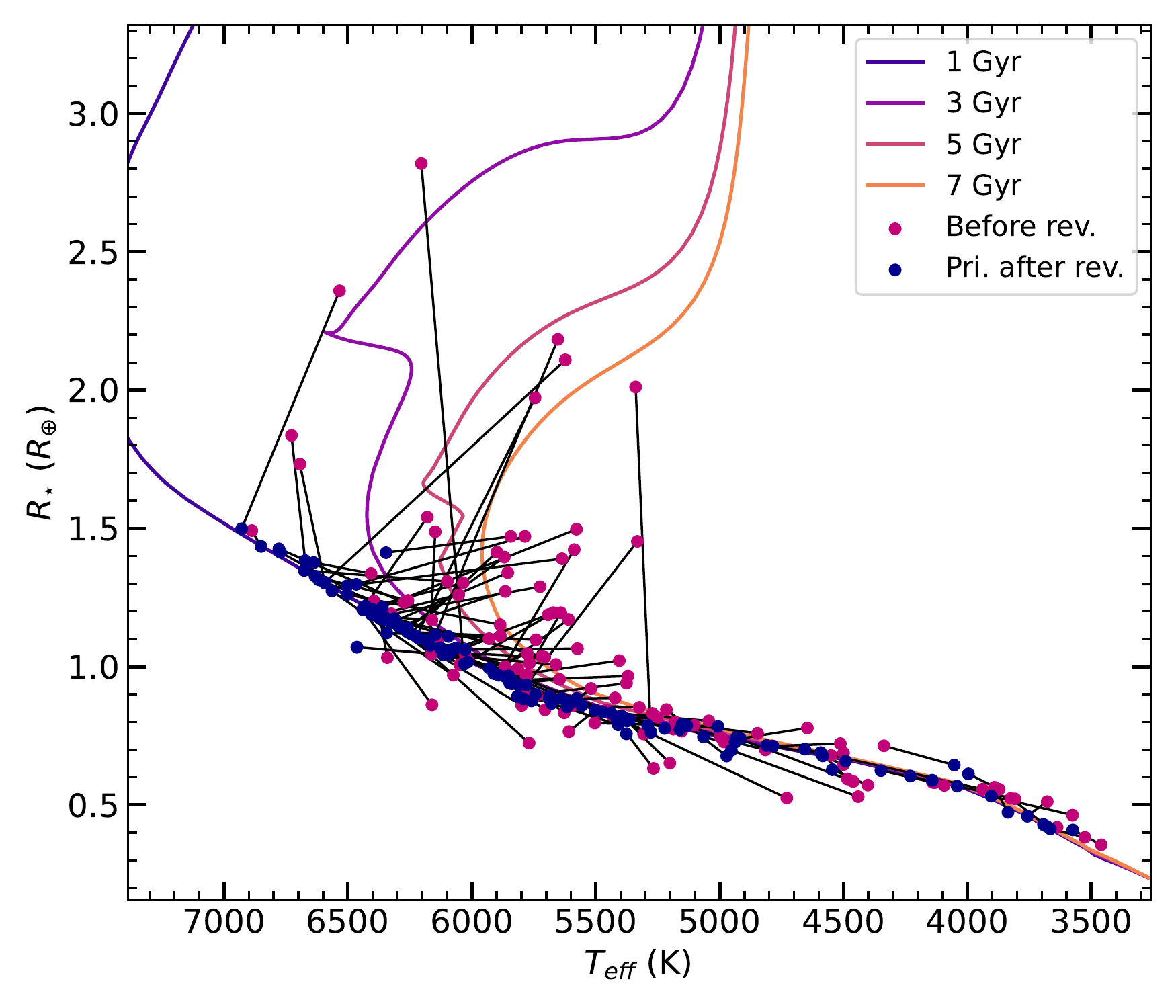}
\caption{Left: A series of histograms of the input composite \citet{Mathur2017} \teff\ (top panel) and the output primary and secondary \teff (middle and bottom panels, respectively). The input temperature distribution and output primary star temperature distribution are both dominated by FGK stars, although there are more hot stars in the retrieved primary star sample because of the upward temperature revision for primaries. Similarly, there are more cool stars and a wide spread for the retrieved secondary star sample because of the downward temperature revision for secondaries. Right: An HR diagram showing the input \citet{Mathur2017} \teff\ and radius for each system (magenta points) and the revised \teff\ and radius for the primary stars (blue points), with lines connecting each system. MIST isochrones at 1, 3, 5, and 7 Gyr are plotted under the points as a reference. We used the 1 Gyr isochrone to calculate the radii, which is a good fit for both the input and output values for most of the stars, although there is scatter in the hottest systems because of significant distance revisions between the KIC and \textit{Gaia} for many of the stars.}
\label{fig:fit_summary}
\end{figure*}

Figure \ref{fig:fit_summary} shows summary plots for the input and revised properties of the stars in our sample. The left panel shows a histogram of the input \kepler\ DR25 \teff\ for each composite system, as well as histograms of the retrieved primary and secondary star temperatures. The reference \kepler\ DR25 values that we compare against throughout this work are from \citet{Mathur2017}, who compiled literature results from photometry, spectroscopy and asteroseismology \citep[e.g.,][]{Huber2014} to produce the stellar properties and distances of the last Kepler planet catalog release (DR25). We were not able to compare to other updated stellar values such as \citet{Berger2020}, because the majority of our binaries were excluded from their analysis. Compared to the composite systems, the revised primary stars have a hotter mode, which was expected given that the primary star temperatures were revised upward. The right panel shows an HR diagram converted to radius-\teff\ space, with magenta points showing the composite system values and blue points showing the output primary star values, wth lines connecting each system. There are MIST isochrones underlying the plot at ages of 1, 3, 5, and 7 Gyr. The cool stars typically have input and output values that fall close to the main sequence. There is more scatter for the hot stars, with many of the input radius values not matching the main sequence. After inspection of these systems, we found that the typical reason for the substantial radius revision was a large change in the measured system distance between our analysis (which typically used \textit{Gaia} distances) and the \citet{Mathur2017} analysis.

We revised the stellar parameters relative to the \kepler\ DR25 parameters \citep{Mathur2017}. On average, the primary star temperatures were 250 K (84th percentile +241 K, 16th percentile -211 K) hotter than the \kepler\ temperature, while the secondary star temperatures were 507 K (84th percentile +988 K, 16th percentile -582 K) cooler than the \kepler\ temperature. In both cases the percentiles quoted are the RMS spread due to the wide range of binary properties. As expected given the demographics of the \kepler\ sample, the majority of our systems had temperatures consistent with FGK spectral types, with a few M dwarfs also present in the sample. This is also visible in the left panel of Figure \ref{fig:fit_summary}.

As part of retrieving the corrected stellar temperature for the primary and secondary stars, we inferred a corrected stellar radius. We calculated a new planetary radius for each planet in the sample by applying a planetary radius correction factor (PRCF; \citealt{Ciardi2015}). The final calculation for the planetary radius, assuming the primary was the host star, was 
\begin{equation}
    \begin{split}
        R_{p, pri} &= R_{p,old} * \frac{R_{\star, pri}}{R_{\star, Kep}} * f_{corr, p}\\
        &= R_{p,old} * \frac{R_{\star, pri}}{R_{\star, Kep}} * \sqrt{1 + 10^{-0.4\Delta m}},
    \end{split}
\end{equation}
where the contrast $\Delta m$ is the contrast in the band where the planet radius was measured, R$_{p, old}$ is the Exoplanet Archive radius, and $R_{\star, Kep}$ is the assumed \kepler\ stellar radius. In this case, we performed synthetic photometry of the binary component spectra in the \kepler\ band and used the resulting contrast when calculating the correction factor. The correction in the case where the secondary star is the planet host is almost identical:
\begin{equation}
    \begin{split}
        R_{p, sec} &= R_{p,old} * \frac{R_{\star, sec}}{R_{\star, Kep}} * f_{corr, s}\\
        &= R_{p,old} * \frac{R_{\star, sec}}{R_{\star, Kep}} * \sqrt{1 + 10^{+0.4\Delta m}}.
    \end{split}
\end{equation}
We calculated separate radii for the individual cases  assuming either the primary or the secondary star was the planet host. We emphasize the results from the primary star host scenario in this work because it is more statistically likely to detect exoplanets around the primary stars (because flux dilution is less severe for a primary star planet host than for the fainter secondary star; \citealt{Gaidos2016}), but future work will assess whether this assumption is appropriate. This will be important because if any of the planets are hosted by the secondary stars, they will constitute a semi-random distribution that dilutes real features in the radius distribution. However, we expect that most of the planets will fall around the primary stars because of detection biases.

\begin{figure*}
\plottwo{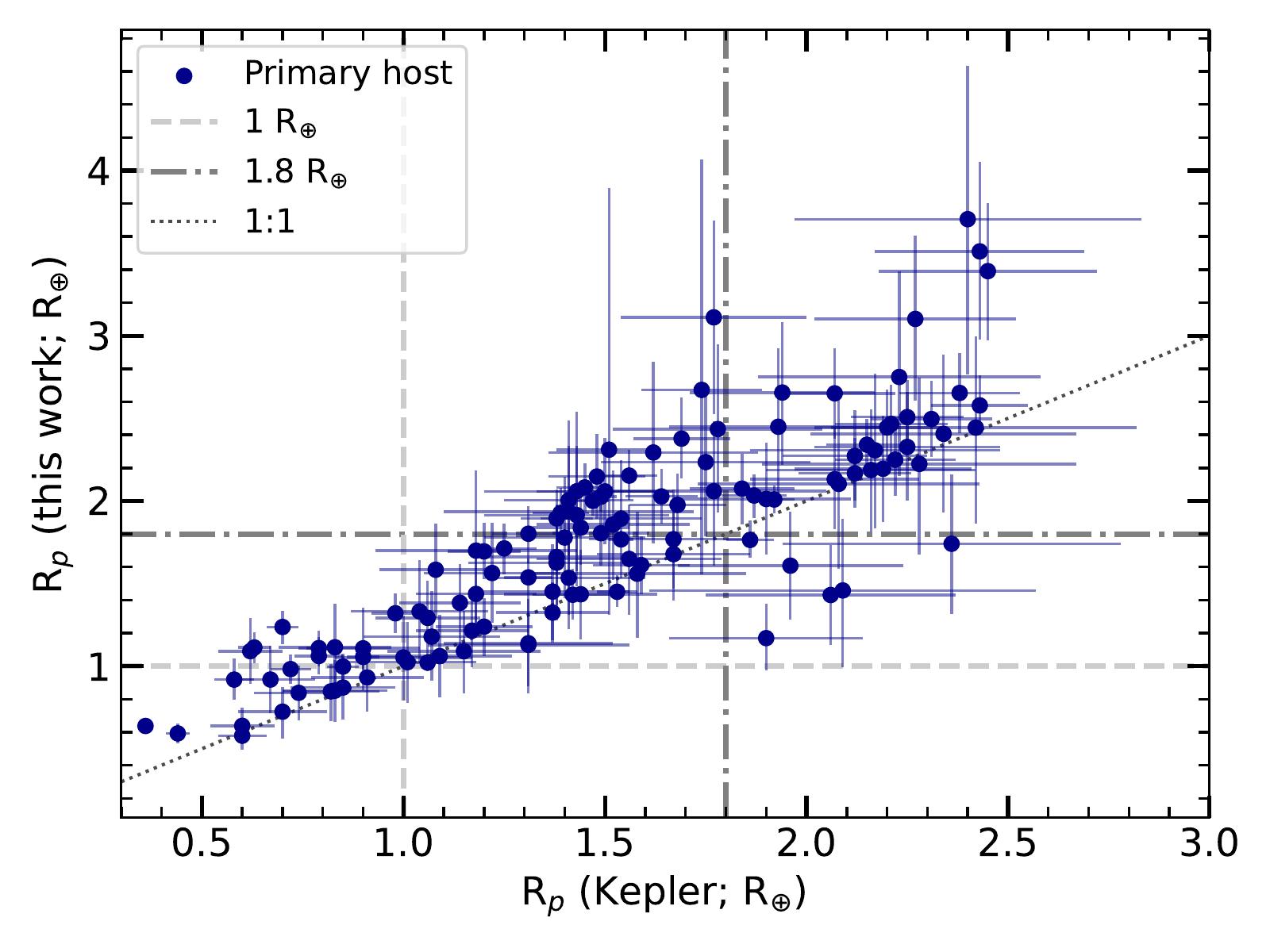}{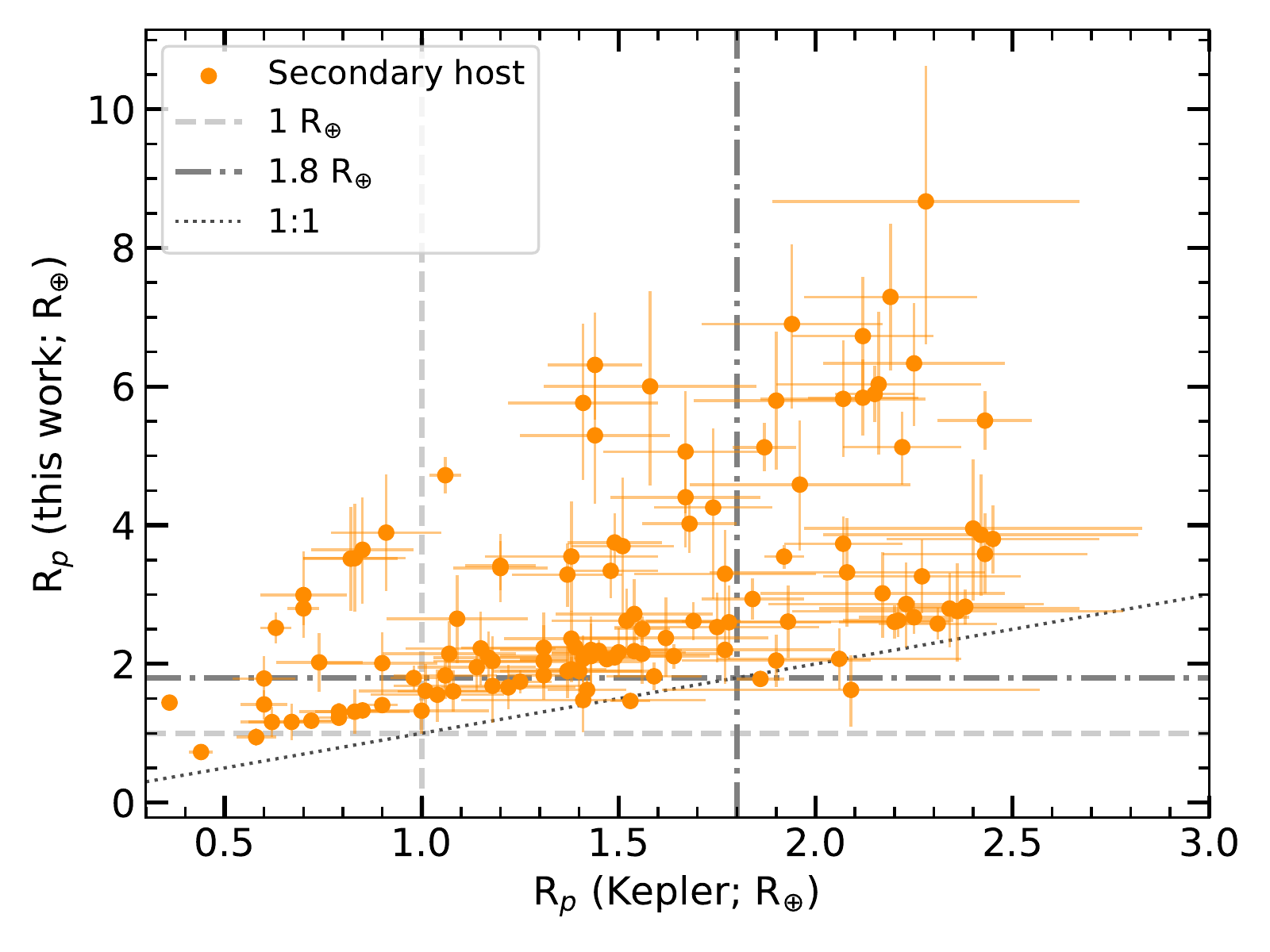}
\caption{Left: Planetary radii from this work, calculated assuming the primary star is the planet host, plotted against the \kepler\ planetary radius. The light gray dashed line denotes 1 \rearth, and the darker gray dashed line denotes 1.8 \rearth\ (approximately the location of the single-star radius gap). The dotted black line denotes one-to-one correspondence between the axes. The majority of systems have larger radii measured in our work than by \kepler\, with planetary radii being revised upward by 17\% (84th percentile +23\%, 16th percentile -16\%), with the large spread in the distribution being caused by the wide range in binary properties within the sample. Right: The same as the left panel, but for the case where the secondary star is assumed to be the planet host. The majority of systems fall above the single-star radius gap, with an average planet radius revision of +66\% (84th percentile +31\%, 16th percentile -127\%).}
\label{fig:fit_compare}
\end{figure*}

To create a final planet sample, we imposed several demographic cuts. Although some systems included planets larger than our 2.5 \rearth\ cutoff, we restricted the sample we analyzed to only include planets with $R_{p} < 2.5$ \rearth, to match the regime where our observations of all sub-Neptunes and super-Earths in known binaries had been taken (i.e., the limits of our original sample selection). We also restricted the sample to planets with periods of $P < 100$ d, and where the original planet radius error was less than 25\%. These cuts were to match the original selection criteria for the CKS sample, and to ensure that we were only analyzing high-quality planetary radius measurements. Our final sample included 122 planets with an average radius error of 13\%, which is comparable to the CKS radius precision. 

Figure \ref{fig:fit_compare} shows the revised planet radii assuming the primary (left) or secondary (right) star is the planet host, plotted against the \citet{Thompson2018} radius. In both cases, the radii from this work are typically larger than the \citet{Thompson2018} radii, with the cases where the secondary star is assumed to be the planet host having larger radius revisions, as expected. The average planetary radius change if the primary star is the host is +17\% (84th percentile +23\%, 16th percentile -16\%), while the average planetary radius change if the secondary star is the host is +66\% (84th percentile +127\%, 16th percentile -31\%). 

\citet{Ciardi2015} introduced a PRCF, as described above, to quantify the planetary radius correction from the presence of a secondary star. We calculated those factors and found that on average the PRCF was typically slightly smaller than the fractional change to the radius that we calculated from the full posterior distributions, and the change was more significant for lower-contrast (higher mass-ratio) binaries. If the primary star was the host, the average PRCF was 1.15$^{+0.25}_{-0.19}$ (or $\sim$15\%), and if the secondary star was the planet host the average PRCF was 1.64$^{+1.19}_{-0.37}$ (or $\sim$65\%). 

The difference between the \citet{Ciardi2015} PRCF and our radius correction is the result of taking the full revised stellar radius into account instead of assuming that the primary star radius is consistent with the \kepler\ radius. The discrepancy between the two values quantifies how much of an impact the radius revision has in addition to the dilution correction. For high-contrast systems, the radius change is typically much less significant than the dilution correction, but for a low-contrast subset (i.e., the systems with the brightest secondary stars), the discrepancy was larger. For some systems, the difference between the PRCF value and the calculated radius revision was up to $\sim50$\%, meaning that the PRCF calculation alone would have under-predicted the planetary radius by 50\%. This suggests that for low-contrast binaries the full stellar characterization is important for achieving accurate planetary parameters, and correcting for the dilution alone is not sufficient.

\section{Revised Planet Radius Distribution}

\begin{figure}
\plotone{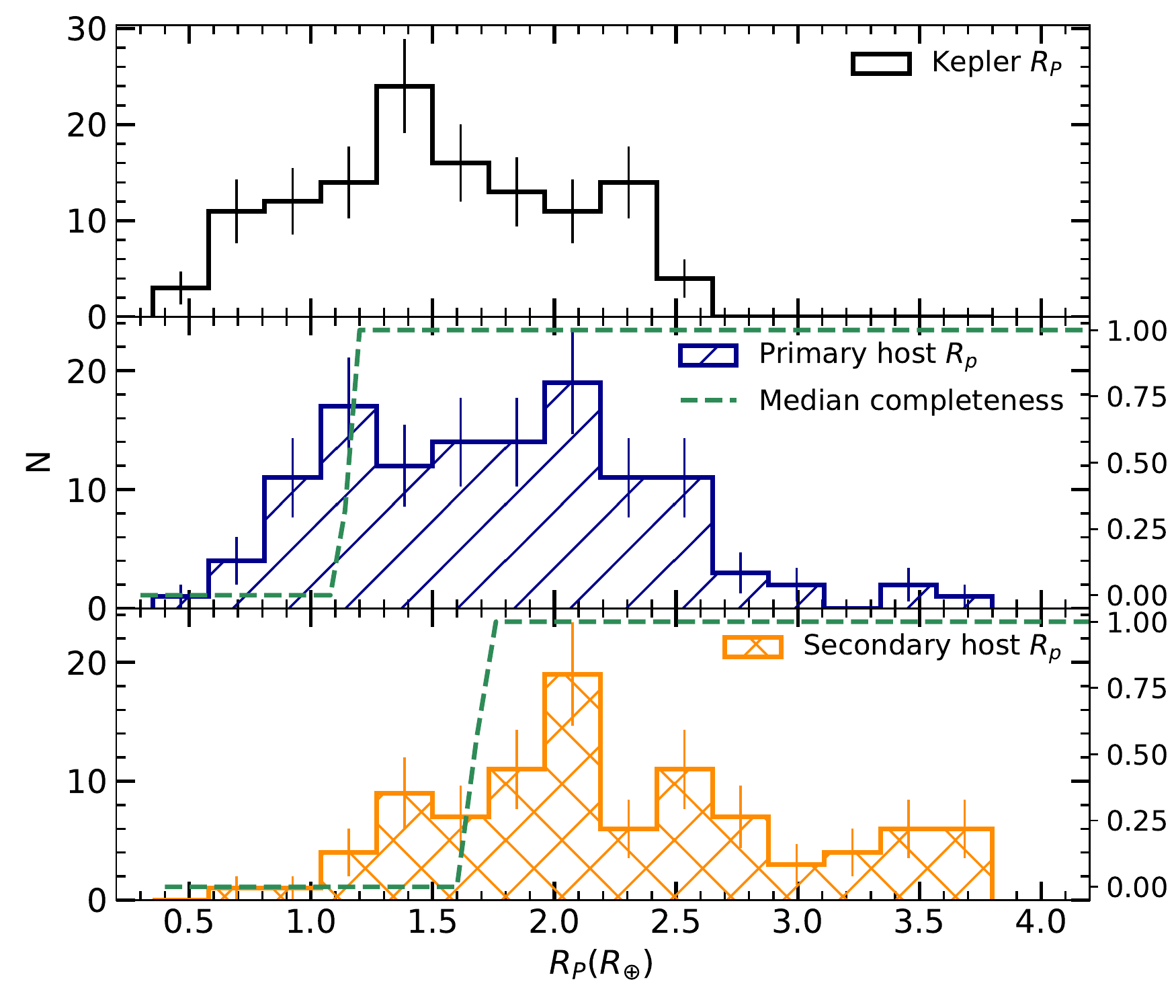}
\caption{Histograms of the input \kepler\ planet radius (top; \citealt{Mathur2017}) and the inferred best-fit planetary radii assuming the primary or secondary stars were the planet hosts (middle and bottom, respectively). The primary and secondary host assumption panels include approximate completeness curves, detailed in the text.}
\label{fig:radius_hist}
\end{figure}

To demonstrate the radius revision for the 122 planets in our sample, Figure \ref{fig:radius_hist} shows the input planetary radius distribution (top panel) and the retrieved planetary radii assuming either the primary or secondary star was the planet host. Statistically, planets are more likely to be found around the primary star, especially for the relatively faint secondary stars, but we present both distributions for reference. 

In Figure \ref{fig:radius_hist}, there are also approximate completeness curves plotted for the observed corrected radius distributions. This back-of-the-envelope calculation was to check that our analysis would have been sensitive to a super-Earth peak or the radius gap, rather than lacking completeness at relatively large planetary radii. We calculated the approximate completeness curves using equation 1 from \citet{Howard2012}: 
\begin{equation}
    S/N = \frac{\delta}{\sigma_{CDPP}} \sqrt{\frac{n_{tr} \cdot t_{dur}}{\text{3 hr}}},
\end{equation}
where $\delta$ is the transit depth, $\sigma_{CDPP}$ is the noise in the transit, $n_{tr}$ is the number of transits, $t_{dur}$ is the transit duration, and the factor of 3 accounts for the duration over which $\sigma_{CDPP}$ was measured. We calculated $t_{dur}$ as
\begin{equation}
    t_{dur} = \frac{P}{\pi}\arcsin{\frac{\sqrt{(R_{p} + R_{\star})^{2} - (bR_{\star})^{2}}}{a}},
\end{equation}
where $P$ is the period, $R_{p}$ and $R_{\star}$ are the planetary and stellar radii, $b$ is the impact parameter, and $a$ is the semimajor axis. To calculate the transit duration we assumed a system with the median properties of our sample: a 1 $M_{\odot}$, 0.96 $R_{\odot}$ star for the primary calculation or a  0.76 $M_{\odot}$, 0.75 $R_{\odot}$ star for the secondary calculation, and a planet with varying radius, a period of 17 days, and an impact parameter of zero. This calculation was not intended to be a rigorous derivation of completeness, but was instead meant to be an approximate calculation to help estimate where completeness was high.

After calculating the transit duration, we calculated a S/N for each simulated planet radius using an estimated transit depth of $\delta = (\frac{R_{p}}{R_{\star}})^{2}$. We assumed the star had the median brightness of our sample ($r' = 13.5$ mag, with $r'$ being approximately analogous to the \kepler\ bandpass), and used Figure 1 of \citet{Jenkins2010} to estimate that $\sigma_{CDPP} = 35$ ppm. For a detection, we required the planet to have $S/N > 7$. We corrected for the presence of the binary star by multiplying the resultant contrast curve's corresponding radii by an approximate average PRCF for either the primary or secondary stars: 1.2 for the primaries and 1.6 for the secondaries. The average primary star dilution factor corresponds to a magnitude difference of $\Delta_{m_{Kep}} = 0.9$ mag, or a mass ratio of $\sim 0.85$ for a typical early G star. Referring to Figure \ref{fig:radius_hist}, the planet sample is approximately complete above 1 \rearth, although this will vary based on the binary star parameters and the properties of each host star. 

\begin{figure*}
\plotone{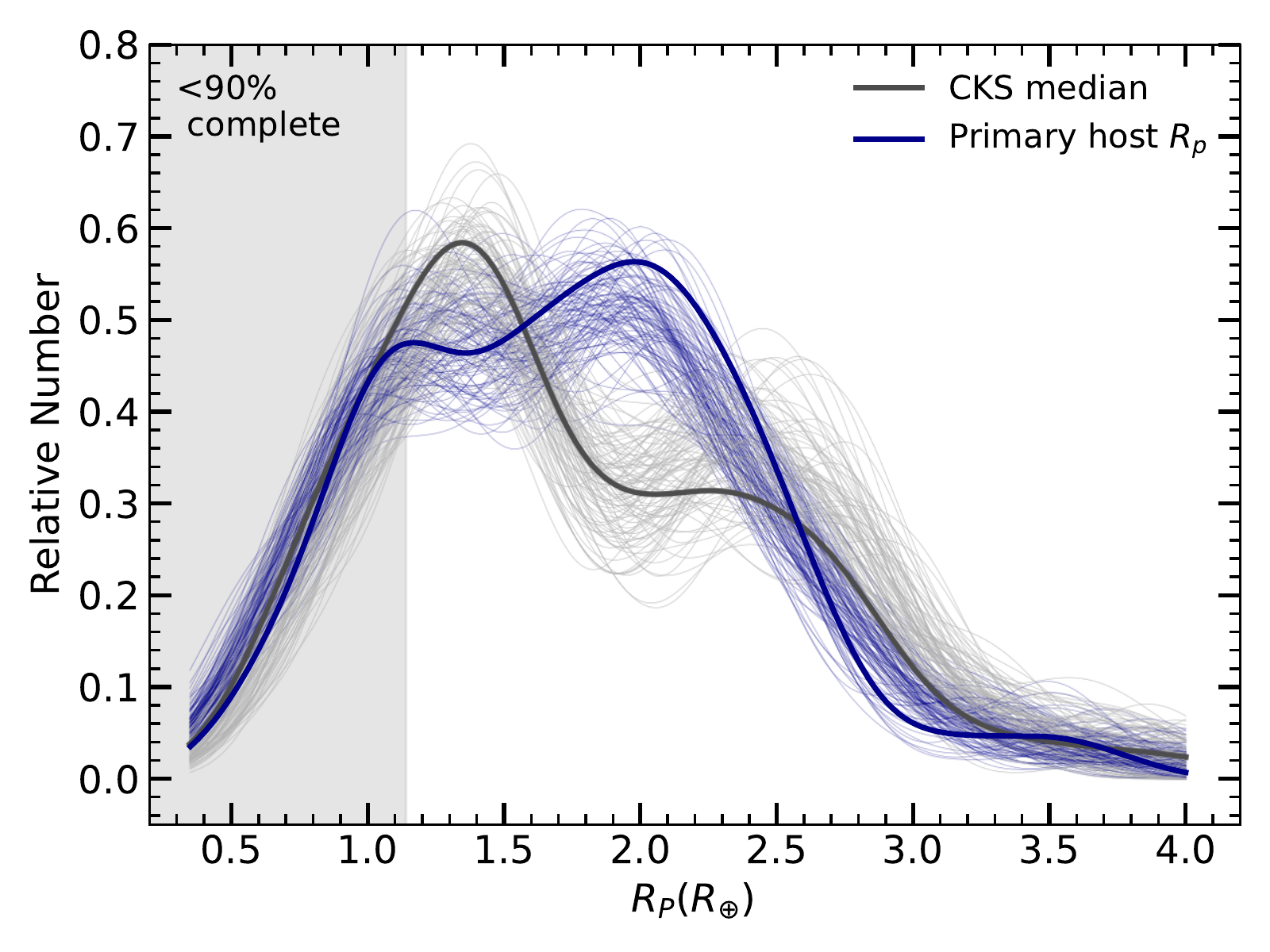}
\caption{A KDE of the revised planetary radii assuming that the primary star is the host in all cases (thick blue line), with KDEs of 100 random draws from the error distributions for each sample underplotted (thin blue lines), as well as the median of the CKS subsample (dark gray line) with KDEs of 100 random subsamples from CKS underplotted (light gray lines). The calculation for the gray region showing the completeness limit is described in the text. Neither the 100 observations of the CKS sample nor the median are consistent with our observed radius distribution.}
\label{fig:radius_hist_pri}
\end{figure*}

To quantitatively compare the binary star planet radius distribution to the CKS radius gap sample of \citet{Petigura2022}, we created a CKS sample with parameters that matched our binary star sample by selecting for planets with $1 \leq R_{p} \leq 2.5$ \rearth\ and periods $P < 100$ d. We performed an Anderson-Darling test on the two distributions after marginalizing over period. An Anderson-Darling test is a statistical test that quantifies whether two samples are drawn from the same underlying distribution. We found that the two distributions are significantly different, with a probability $p < 0.05$ that the samples are drawn from the same population. 

As another comparison between the single-star sample and our results, we drew 100 random subsamples of the matching CKS sample with a sample size equal to the binary star sample. Each of these subsamples was equivalent to an observation of the single star radius gap using our sample size, and was intended as a quantification of the scatter that would be expected to occur if the radius gap exists but the sample size is small. As Figure \ref{fig:radius_hist_pri} demonstrates, the single-star radius gap remains resolved in the CKS subsamples, and the random draws (as well as the median) are significantly discrepant with our observed population. If the radius gap in binary stars was identical to the gap observed in planets in single stars, we would expect that our observed planetary radius distribution would be consistent with at least some of the distributions drawn from the ``observed'' CKS subsamples. Instead, the binary sample appears to have an underabundance of super-Earths and too many objects in or near the radius gap.

In single stars, the radius gap appears at $\sim 1.75-1.8$ \rearth. We do not observe a corresponding gap in the binary star distribution. The gap in \citet{Fulton2017} is well-defined over several bins, while our distribution is uniform within statistical uncertainty. Therefore, we conclude that we do not observe a gap in the radius distribution for binary stars.

\section{The Absence of a Radius Gap in Binary Stars}
To explore the radius distribution of small planets in binary stars, we observed a sample of 119 close and intermediate separation ($\rho < 2\arcsec$) binaries hosting 122 planets. Using new stellar radii, we revised the planet radii, and found that we do not observe a statistically significant radius gap for planets in binary stars. In the following section, we discuss the implications of the radius distribution of planets in binary stars for the formation of planets in binary star systems.

\begin{figure*}
\plotone{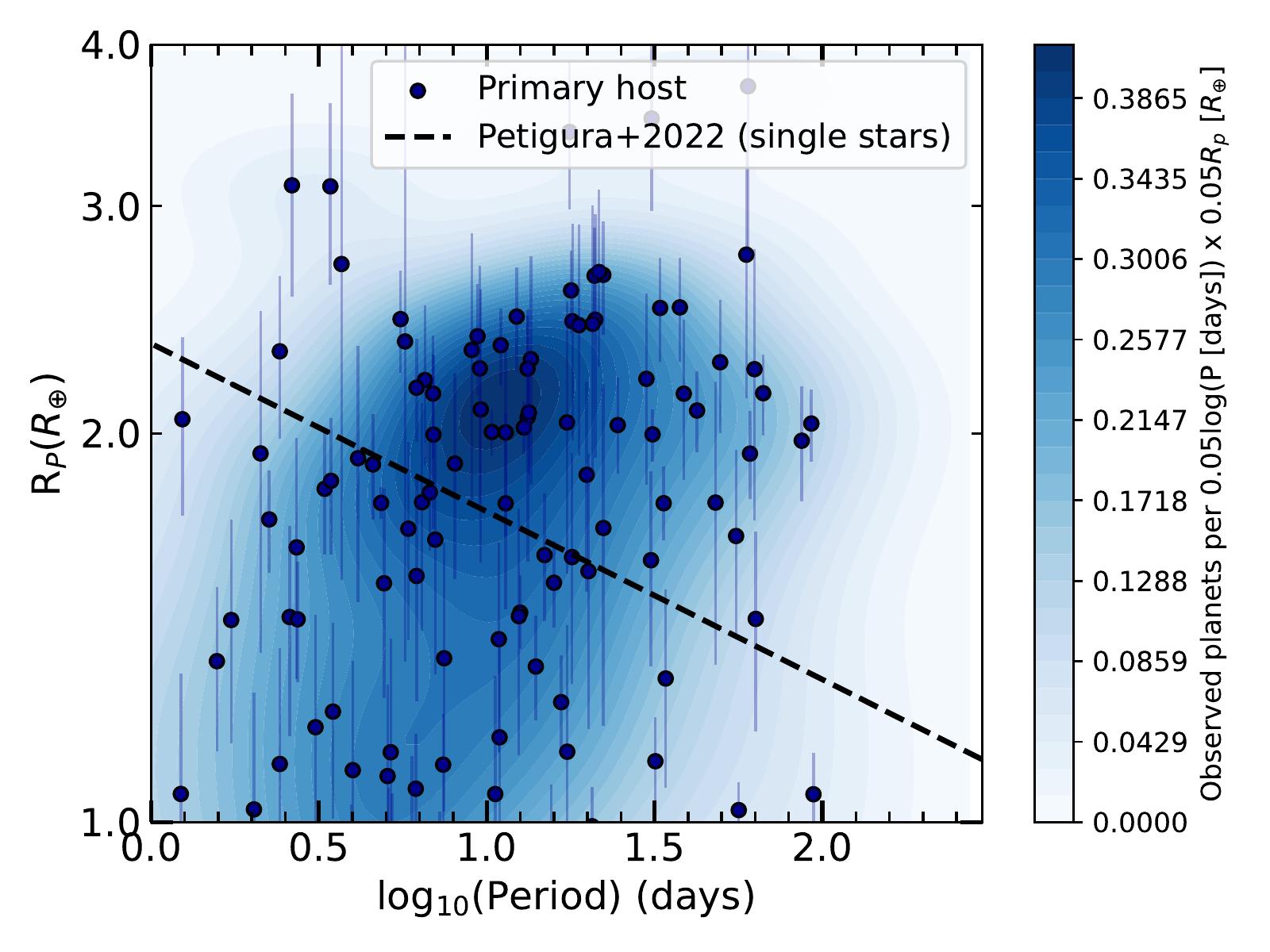}
\caption{The planet radius assuming the planets all orbit the primary stars, plotted against the planets' orbital periods (blue points), with a kernel density estimate (KDE) underlaid (blue contours). The KDE has a kernel size of 0.25 in both log(period) and linear radius and shows no clear gap in the period-radius distribution. All quantities are in terms of detected planets, not occurrence rate.}
\label{fig:radius_gap}
\end{figure*}

Figure \ref{fig:radius_gap} shows another visualization of the planetary radius gap, plotting planet radius as a function of the planetary orbital period. Underlying the scatter plot is a kernel density estimate (KDE) of the scatter plot, which we calculated using a two-dimensional Gaussian kernel with a bandwidth of 0.25 in both log(period) and radius space. The KDE bandwidth was determined by optimizing to find the best-fit KDE for the observed distribution, to remove human bias in the choice of kernel size. We assumed all planets orbited the primary star because it is more statistically likely for a planet to be discovered around a brighter primary star than a fainter secondary star. We expect that if a few planets do orbit around the secondary star that they would add random scatter without significantly changing our conclusions. If a significant fraction of these planets orbit around the secondary star, it may limit our sensitivity to a radius gap among binary stars, but this scenario is unlikely.

Figures \ref{fig:radius_hist_pri} and \ref{fig:radius_gap} show no clear bimodality. Figure \ref{fig:radius_gap} shows a peak around the location of the sub-Neptune occurrence peak ($\sim 2.3$ \rearth), but there is no corresponding increase in the number of super-Earths at smaller radii, even though there should have been adequate survey sensitivity to detect planets down to $\sim$1 \rearth, even after accounting for the binary flux dilution. There may be a slight overdensity of planets around $\sim$1.2 \rearth, but we do not have a sufficiently large sample to assert that we observe a second peak at small radii.

The single star radius gap occurs at 1.75-1.9 \rearth\ \citep[e.g.,][]{Hardegree-Ullman2020, Petigura2022}, and is thought to be the result of atmospheric mass loss, meaning that the population of super-Earths is likely a mixture of the high-mass end of terrestrial planet formation and the stripped cores of former sub-Neptunes \citep[e.g.,][]{Rogers2021}. The single star radius gap is mass-, metallicity-, and age-dependent \citep{Petigura2022}, with a slope and mean location that vary with all three parameters. However, there are no single-star populations in the CKS sample that do not show evidence for a radius gap: systems with a variety of metallicities, stellar masses, and ages still demonstrate gaps, just with different locations and slopes.

One possible explanation for not observing the radius gap in our sample would be that the errors on our radii are too large to resolve a gap. However, we removed any planets with an initial radius error $\sigma_{R} > 25\%$, and our average planetary radius error was $\sim 13$\%, which is comparable to the typical CKS precision of $\sim$ 10\% \citep[e.g.,][]{Petigura2022}. Therefore, we do not expect that large radius errors are significantly impacting our results. The gap is resolved in other samples beyond the \kepler\ sample \citep[e.g.,][]{Hardegree-Ullman2020, Rains2021}, suggesting that it is not a feature of the \kepler\ sample alone. Thus, we also do not expect the lack of a gap to be caused by selecting a different sample than CKS. 

Another possibility is that there are not enough systems to resolve a gap. The KDE in Figure \ref{fig:radius_gap} shows hints of an elongation toward shorter periods and smaller radii, which is similar to the period-radius offset between the radius peaks for super-Earths and sub-Neptunes. It is possible that we have too few super-Earths to resolve a peak at small radii, and thus to resolve a gap. Although the input radius distribution does appear to marginally resolve a gap (top panel of Figure \ref{fig:radius_hist}), the population of super-Earths is substantially decreased because of the radius revision, which could remove the apparent bimodality after radius revision. However, as Figure \ref{fig:radius_hist_pri} demonstrates, random subsamples drawn from the \citet{Petigura2022} sample with an identical number of planets and comparable radius precision do resolve a gap, suggesting that sample size is not a factor in our nondetection.

Stellar multiplicity could also physically change the properties of the radius valley. Binary properties are typically separation-dependent, including disk lifetime \citep{Cieza2009, Kraus2012}, disk size \citep{Harris2012, Zurlo2020}, and planet occurrence \citep{Kraus2016, Moe2021}. Any changes to exoplanet properties caused by binaries is likely also separation dependent, and may also be mass ratio dependent: a more massive secondary star may be more gravitationally disruptive or remove more material from the mass reservoir of the system during formation. Some of these impacts of stellar multiplicity may alter the mature observed planet population by impacting the core mass distribution of the forming planets.

One possible change to the properties of exoplanets in binary stars is a change in the location of the peak of the core mass distribution as a function of binary separation. Binaries truncate disks \citep{Harris2012}, reduce disk lifetimes \citep[e.g.,][]{Cieza2009, Kraus2012}, and decrease the planet occurrence rate \citep[e.g.,][]{Wang2015, Kraus2016, Moe2021}. These effects are most significant in close binaries (separations $\rho < 100$ au), but persist to wider separations of up to 1000 au, suggesting that it could be harder to form more massive cores in most binaries. \citet{Harris2012} found that the mm flux from binary disks in Taurus decreases as a function of binary separation, indicating that there may be less dust in close binary disks, which could suppress planet formation. \citet{Barenfeld2019} compared the population of disks in Upper Sco (10 Myr; \citealt{Pecaut2012, Feiden2016, Sullivan2021}) and Taurus (1-2 Myr; \citealt{Krolikowski2021}) and found that the mm flux from disks in binaries at young ages is similar to that of older disks in single stars. This result suggests that similar processes are occurring to reduce mm flux in binaries and single stars, but at different speeds, potentially reducing the timescale and mass reservoir for planet formation in binaries. However, \citet{Barenfeld2019} also noted that there are several mechanisms for altering mm flux from disks, including optical depth effects and dust fragmentation, suggesting that the differences between binary and single star disks may not just be a result of differing disk masses. 

These studies exploring circumstellar disks in binary systems suggest that the planet-forming environemnt in binaries resembles single-star planet formation, just on shorter timescales and smaller disks. These factors could inhibit core formation by causing planet formation to cease earlier in binaries than in single stars, or by providing a smaller mass reservoir to begin with. We would expect this effect to be separation-dependent, reflecting the separation dependence of disk lifetime and disk size.

The average stellar separation of our sample is $\sim$500 au, which is an order of magnitude larger than the regime where the most dramatic suppression of disks and planets occurs in binaries. However, binaries with separations of hundreds of au still have suppressed disk mm emission relative to single stars \citep{Harris2012}, suggesting that the impacts of multiplicity persist at separations well beyond those of the close binaries. \citet{Wang2014} also suggested that planet occurrence is suppressed in binaries with separations as wide as 1500 au, supporting the conclusion that although very close binaries have the most detrimental effect on planet survival, wider binaries likely still impact planet formation. However, whether reduced disk masses and lifetimes should result in a different core mass distribution or simply fewer cores still drawn from a universal core mass distribution is unclear. If the core mass distribution does change as a function of binary separation and mass ratio, then any gap present in the data should change location, and the range of binary properties covered by our sample would blur out the gap, producing fewer systems systematically over a wide range of planetary radii rather than a single defined gap.

One way to probe the possible effect of binary separation and mass ratio on the location of the radius gap would be to have a large enough sample to bin by mass ratio and/or binary separation. Then, we could see if we observed a gap at progressively larger planetary radii, with the location of the binary star gap approaching the location of the single star gap as mass ratio decreased and binary separation increased. However, our sample of $\sim$ 100 planets is not large enough to bin. Future work should explore other avenues for expanding the sample of small planets ($R_{p} < 1.8$ \rearth) observed in binary stars, either with higher sensitivity observations from ground-based observatories or TESS observations with long time baselines (e.g., from the continuous viewing zones of TESS) combined with multiplicity follow-up. In general, for future targeted observations of planets in binaries to explore the demographics, a larger sample of planets in binary stars potentially collected using a sample of binaries to be observed by PLATO or other upcoming transiting exoplanet discovery and characterization missions is needed to be able to explore planet properties without marginalizing over binary properties like mass ratio and separation.

\section{Conclusions}
To explore the properties of the radius gap in binary stars, we have observed a large sample of binary stars that host small planets ($R_{p} < 2.5\ R_{\oplus}$) using the HET. We reanalyzed the stars to retrieve the parameters of the individual components, then used the revised stellar temperatures and radii to obtain corrected planetary radii. 

We did not detect a radius gap in the sample of planets in binary systems. This nondetection should be probed with different and larger samples to be confirmed, but could be caused by changes in the core mass function of planets induced by the presence of the secondary star, which could create a gap with a location that is dependent on binary separation and system mass ratio. This could blur out the location of a gap in our sample, which has a wide range of physical separations and mass ratios, reducing our ability to detect a gap when marginalizing over separation and secondary mass. Future work exploring the separation and mass ratio dependence of the radius gap in binary stars should further illuminate the role of the secondary star in shaping the radius gap, which has implications for planet formation theory in both single and binary stars. To perform this future work, a larger sample of planets in binaries with accurately and precisely measured parameters will be vital, and can be provided with a sample of binaries observed by upcoming missions like PLATO or Earth 2.0.

\software{emcee \citep{Foreman-Mackey2013}, corner \citep{Foreman-Mackey2016}, scipy \citep{Virtanen2020}, astropy \citep{astropy2013, astropy2018, astropy2022}, numpy \citep{Harris2020}, matplotlib \citep{Hunter2007}, pyphot\footnote{\url{https://mfouesneau.github.io/pyphot}}, dustmaps \citep{Green2018}, scikit-learn \citep{scikit-learn}}\\

K.S. acknowledges that this material is based upon work supported by the National Science Foundation Graduate Research Fellowship under grant No. DGE-1610403. D.H. acknowledges support from the Alfred P. Sloan Foundation and the National Aeronautics and Space Administration (80NSSC19K0597). E.P. acknowledges support from the Alfred P. Sloan Foundation. T.A.B.’s research was supported by an appointment to the NASA Postdoctoral Program at the NASA Goddard Space Flight Center, administered by Oak Ridge Associated Universities under contract with NASA. The authors sincerely thank the observing staff and resident astronomers at the Hobby-Eberly Telescope for obtaining the observations presented in this work. The authors acknowledge the Texas Advanced Computing Center (TACC) at The University of Texas at Austin for providing high-performance computing resources that have contributed to the research results reported within this paper.

The Hobby-Eberly Telescope (HET) is a joint project of the University of Texas at Austin, the Pennsylvania State University, Ludwig-Maximilians-Universität München, and Georg-August-Universität Göttingen. The HET is named in honor of its principal benefactors, William P. Hobby and Robert E. Eberly. The Low-Resolution Spectrograph 2 (LRS2) was developed and funded by the University of Texas at Austin McDonald Observatory and Department of Astronomy and by Pennsylvania State University. We thank the Leibniz-Institut für Astrophysik Potsdam (AIP) and the Institut für Astrophysik Göttingen (IAG) for their contributions to the construction of the integral field units. 

Some of the data presented herein were obtained at the W. M. Keck Observatory, which is operated as a scientific partnership among the California Institute of Technology, the University of California and the National Aeronautics and Space Administration. The Observatory was made possible by the generous financial support of the W. M. Keck Foundation. The authors wish to recognize and acknowledge the very significant cultural role and reverence that the summit of Maunakea has always had within the indigenous Hawaiian community.  We are most fortunate to have the opportunity to conduct observations from this mountain. 

This publication makes use of data products from the Two Micron All Sky Survey, which is a joint project of the University of Massachusetts and the Infrared Processing and Analysis Center/California Institute of Technology, funded by the National Aeronautics and Space Administration and the National Science Foundation. This research has made use of the SVO Filter Profile Service (\url{http://svo2.cab.inta-csic.es/theory/fps/}) supported from the Spanish MINECO through grant AYA2017-84089. This research has made use of the VizieR catalogue access tool, CDS, Strasbourg, France (DOI : 10.26093/cds/vizier). The original description of the VizieR service was published in 2000, A\&AS 143, 23. This work has made use of data from the European Space Agency (ESA) mission {\it Gaia} (\url{https://www.cosmos.esa.int/gaia}), processed by the {\it Gaia} Data Processing and Analysis Consortium (DPAC, \url{https://www.cosmos.esa.int/web/gaia/dpac/consortium}). Funding for the DPAC has been provided by national institutions, in particular the institutions participating in the {\it Gaia} Multilateral Agreement. This research has made use of the Exoplanet Follow-up Observation Program website, which is operated by the California Institute of Technology, under contract with the National Aeronautics and Space Administration under the Exoplanet Exploration Program. This research has made use of the NASA Exoplanet Archive, which is operated by the California Institute of Technology, under contract with the National Aeronautics and Space Administration under the Exoplanet Exploration Program.

\begin{longrotatetable} %
\begin{deluxetable*}{ccCCCCccCCCCCCC}
\setcounter{table}{1}
\tablecaption{System Parameters for Each Source \label{tab:source_params}}
\tablecolumns{13}
\tablehead{
\colhead{KOI} & \colhead{$\rho$} & \colhead{Obs date} & \colhead{r'} & \colhead{S/N} & \colhead{$\Delta m_{i}$} & \colhead{$\Delta m_{LP600}$} & \colhead{$\Delta m_{Gaia}$} & \colhead{$\Delta m_{562 nm}$} 
& \colhead{$\Delta m_{692 nm}$} & \colhead{$\Delta m_{880 nm}$} & \colhead{$\Delta m_{J}$} & \colhead{$\Delta m_{K}$}\\
\colhead{} & \colhead{('')} & \colhead{} & \colhead{(mag)} & \colhead{} & \colhead{(mag)} & \colhead{(mag)} & \colhead{(mag)} & \colhead{(mag)} 
& \colhead{(mag)} & \colhead{(mag)} & \colhead{(mag)} & \colhead{(mag)}\\
}
\startdata  
0042 & 1.66 & 20220713 & 9.30 & 2095.0 & \nodata & 3.04 $\pm$ 0.17 & \nodata & 4.24 $\pm$ 0.15 & \nodata & \nodata & 2.212 $\pm$ 0.026 & 1.873 $\pm$ 0.024 \\
0177 & 0.232 & 20210905 & 12.93 & 227.0 & 0.97 $\pm$ 0.35 & \nodata & \nodata & \nodata & 0.821 $\pm$ 0.198 & 0.613 $\pm$ 0.233 & \nodata & 0.185 $\pm$ 0.01 \\
0191 & 1.672 & 20221020 & 14.68 & 136.0 & 2.746 $\pm$ 0.023 & 3.09 $\pm$ 0.49 & \nodata & \nodata & \nodata & \nodata & \nodata & \nodata \\
0227 & 0.311 & 20220803 & 13.45 & 237.0 & \nodata & 0.84 $\pm$ 0.09 & \nodata & 1.03 $\pm$ 0.15 & 1.5 $\pm$ 0.15 & \nodata & 0.018 $\pm$ 0.01 \\
0270 & 0.145 & 20221001 & 11.031 & 582.0 & \nodata & \nodata & \nodata & \nodata & 0.605 $\pm$ 0.15 & 0.768 $\pm$ 0.15 & 0.0 $\pm$ 0.02 & 0.447 $\pm$ 0.322 \\
\enddata
\tablecomments{The full table is available in machine-readable format online.}
\end{deluxetable*}
\end{longrotatetable}

\bibliography{mcmc3_bib}
\end{document}